\documentclass[12pt,preprint]{aastex}

\slugcomment{Submitted, \today}

\shorttitle{Gemini Adaptive Optics on 3C 230}
\shortauthors{Steinbring}

\input epsf
\def\plotone#1{\centering \leavevmode
\epsfxsize=1.0\columnwidth \epsfbox{#1}}

\def\plotonenarrow#1{\centering \leavevmode
\epsfxsize=0.667\columnwidth \epsfbox{#1}}

\begin{document}

\title{Radio Galaxy 3C 230 Observed with
Gemini Laser-Adaptive-Optics Integral-Field Spectroscopy}

\author{Eric Steinbring\altaffilmark{1}}

\altaffiltext{1}{Herzberg Institute of Astrophysics, National Research Council
Canada, Victoria, BC V9E 2E7, Canada}

\begin{abstract}

The Altair laser-guide-star adaptive optics facility combined with the
Near-Infrared Integral Field Spectrometer (NIFS) on Gemini North have been
employed to study the morphology and kinematics of 3C 230 at $z=1.5$, the first such
observations of a high-redshift radio galaxy. These suggest a bi-polar 
outflow spanning 0\farcs9 ($\sim 16$~kpc projected distance for a standard 
$\Lambda$ CDM cosmology) reaching a mean relative velocity of $235~{\rm km}~{\rm s}^{-1}$ in
redshifted H$\alpha+$[N {\small II}] and [S {\small II}] emission.
Structure is resolved to 0\farcs1 (0.8 kpc), well correlated with optical images
from the Hubble Space Telescope and Very Large Array radio maps obtained at
similar spatial resolution. Line diagnostics suggest that over the $10^7$~yr to
$10^8$~yr duration of its AGN activity, gas has been ejected into bright
turbulent lobes at rates comparable to star formation, although constituting
perhaps only 1\% of the baryonic mass in the galaxy.

\end{abstract}


\section{Introduction}\label{introduction}

Feedback from AGNs is favoured in current models of 
galaxy formation and evolution to explain the flattened upper end of the galaxy
mass spectrum \citep[e.g.][]{Silk1998, Benson2003, DiMatteo2005, Bower2006, Croton2006}.  
A period of rapid star formation is truncated by the AGN heating or removing cold gas
via winds, leaving massive galaxies to passively evolve.
Evidence of AGN-driven outflows is found in high-redshift radio galaxies (HzRGs), close to the epoch when the bulk of their stars formed. Initial long-slit optical spectrosopy revealed $1<z<2$ HzRGs surrounded by 
halos of broad emission-line gas over 10-100 kpc
scales \citep[e.g.][]{McCarthy1996, VillarMartin1999, VillarMartin2003}, although the
kinematics of the outer parts of these halos may in some cases be quiescent,
rotating, or perhaps even in-falling \citep[e.g. MRC
2104-242;][]{VillarMartin2006}. But recent work aided with integral-field
near-infrared (NIR) spectrographs, have shown high-velocity ($\sim1000~{\rm km}
{\rm s}^{-1}$) outflows at kpc scales for $z\sim2$ and higher redshift RGs
\citep{Nesvadba2006, Reuland2007, Nesvadba2008}. These HzRGs are witnessed in
the throes of ejecting large fractions of their gas ($>10^{10} M_\odot$)
via AGN-driven winds, based on their H$\alpha$ luminosities, estimated timescale
of AGN activity, and ionization conditions inferred from diagnostic
emission-line ratios. A disadvantage of seeing-limited studies, however, is that
they lack the spatial resolution needed to study HzRG substructure on sub-kpc
scales (1\arcsec~$\approx8$ kpc at $z=2$ for a $H_0=70$ km ${\rm s}^{-1}$ Mpc,
$\Omega_{\rm m}=0.3$, $\Omega_{\Lambda}=0.7$ concordance cosmology, which is
adopted throughout). Both HST and ground-based adaptive-optics (AO) imaging
shows them to be typically a string of compact knots with separations
of a few kpc, aligned with the radio axis, and with colors consistent with a
young stellar population \citep[e.g.][]{Steinbring2002, Stockton2004, Zirm2005}.
Probing the kinematics and ionization conditions on these scales could
decisively find correlations between the radio core, jet, and gas outflow;
integral field-units behind AO being a promising new tool.

3C 230 \citep[$z=1.487$,][]{Hewitt1991} is an excellent HzRG for such a study: 
it is a powerful radio source \citep[19.2 Jy at 178 MHz;][]{Kellerman1969} and
particularly bright in the near infrared \citep[$K=17.3$;][]{Steinbring2002}. It
is at the bright edge of the observed $K-z$ relation \citep{Willott2003}, 
near a galaxy mass-limit
of $10^{12}~M_\odot$ predicted by models \citep[e.g.][]{RoccaVolmerange2004}. A star ideal 
for AO guiding is in nearby projection on the sky, and was used to
assist $H$ and $K$-band imaging on CFHT \citep{Steinbring2002}. Those ${\rm
FWHM}=$~0\farcs24 resolution images were a good match to archival HST WFPC2
F702W optical images and ${\rm FWHM}\approx$~0\farcs4 VLA radio maps at 4.8 and
8.4 GHz \citep{Rhee1996}. It is a wide double-lobed RG, with a greatest angular 
extent of 5\farcs9 or approximately 51 kpc in projection. For a typical jet-head 
velocity of 0.1c \citep[within the range 0.01c to 0.2c;][]{Wellman1997}, 
AGN activity dates back at least $2\times10^7$ yr, but with de-projection and 
lower jet speed this could be an order of magnitude higher, up to a few 
$\times10^8$ yr. The F702W and $H$ images discussed by Steinbring et al. 
are aligned with the radio axis, comprising three 
marginally resolved sub-components with total $M_H\approx-26$. Even so, 
the overall $R-H=3.5$ color (restframe $\approx U-R$) is consistent with a stellar 
population less than 1 Gyr old.  The $K$ image is more compact than 
$H$, which could be explained by dust obscuration, with an $H-K=0.4$ color 
also consistent with a young population - implying that most star formation in 3C 
230 was intense and brief.

This paper discusses AO observations of 3C 230 obtained with the Near-Infrared 
Integral Field Spectrometer (NIFS) combined with the Altair laser facility on 
Gemini North, a first attempt at providing kinematic and line diagnostic 
information at the same spatial resolution as previous UV continuum imaging with 
HST and radio maps from the VLA \citep{Steinbring2010}. These new results - the 
first laser-AO assisted integral-field spectroscopy on a HzRG - are presented in 
Section~\ref{data} and analyzed along with the available archival data. An 
inherent disadvantage of studying high-$z$ galaxies with AO, generally with 
small spatial pixels to sample a diffraction-limited point-spread function 
(PSF), is poor sensitivity to faint nebulous emission. Here the focus is on 
bright compact features (where AO is most sensitive) and correlating their 
kinematics with the radio, along with looking for any trend in diagnostic 
line ratios. Discussion follows in Section~\ref{discussion}, pointing to a case 
of possible AGN-driven outflow at the core of 3C 230.

\section{Archival Data, Observations, Reductions and Combined 
Analyses}\label{data}

Archival data for 3C 230 are shown in Figure~\ref{plot_finder}; the HST WFPC2 
F702W image is overplotted with contours from CFHT $K$-band AO data. The 
previous AO imaging was facilitated by the relatively bright star ($R=14.7$ mag) 
USNO 0825-07044237 at ${\rm R.A}=$~09h51m58.554s ${\rm 
Dec.}=$~-00$^\circ$01\arcmin28\farcs19 (J2000.0), 4\arcsec~to the west of the 
galaxy. Comparable resolution VLA A-array maps at 8.4 GHz have since become 
available \citep[data kindly provided by Teddy Cheung]{deKoff2000} and are 
overplotted. These radio maps have improved resolution ($\sim$0\farcs1) over 
previous ones and are deeper, detecting the radio core at ${\rm 
R.A.}=$~9h51m58.90s and ${\rm Dec}=$~-00$^\circ$01\arcmin27\farcs87 (J2000.0), 
which coincides with the peak of emission in K observed with CFHT. Another, 
slightly fainter, radio peak 0\farcs18 away 
(9h51m58.91,-00$^\circ$01\arcmin28\farcs11) will be discussed in more detail to 
follow.

\subsection{Gemini Laser-Adaptive-Optics Observations}

Gemini North observations were obtained on the nights of 23 and 24 April 2008
with the Near-infrared Integral Field Spectrometer \citep[NIFS,][]{McGregor2003}
behind the Altair AO facility in laser-guide-star (LGS) mode \citep{Boccas2006}.
During LGS operation, a natural guide star (NGS) is still needed to sense tip
and tilt phase errors, although it can be fainter than necessary for full NGS
correction ($R\approx12$). For 3C 230, the nearby star was ideal for tip-tilt
sensing by Altair. Reliable astrometry provided by the previous CFHT AO
observations was particularly helpful, as the target faintness and small NIFS
field of view (3\arcsec $\times$ 3\arcsec) required offsets to be made ``blind."
Individual exposures were 600 s, obtained in a box-dither pattern of 0\farcs2
diameter, and interleaved with offsets to blank sky, for a total of 6600 s
exposure on target. The sky offset was 3\arcsec~ (to the north of the target) in
a sky-target-target-sky pattern. For these small dithers and sky offsets no
re-acquisition is necessary, and so overheads are no more than for NGS mode.
Non-AO corrected images of standard stars taken immediately before or after
observations on both nights indicate near-median seeing conditions 
(${\rm FWHM}<0\farcs7$~at $V$). From previous experience with Altair this is known
to translate into stable diffraction-limited performance ($H$-band FWHM of 0\farcs1
or better) with Strehl ratio $S$ - peak intensity relative to that for an
ideal diffraction pattern - close to 20\%.  Anisokinetic error (tip-tilt
anisoplanatism) is expected to be negligible due to the small
target-to-guidestar offset \citep{Steinbring2005}.

The NIFS $H$ grating and $JH$ filter were used, yielding a spectral resolution
of $\lambda/{\delta \lambda}\approx5500$ and full-width at half maximum (FWHM)
velocity resolution of $\sim60$ km ${\rm s}^{-1}$. For the faint spectral
features observed here, excellent sky subtraction was critical to achieve useful
$S/N$. For a few of the observations no associated sky image was available, due
to the telescope being called away for a target-of-opportunity observations.
Those observations without an associated sky have been excluded from analysis.
The sky-subtracted observations were stacked to produce the data cube using a
combination of standard IRAF packages and custom Interactive Data Language (IDL) 
utilities. Spectra were
wavelength calibrated using a spectrum of an arc-lamp, with telluric and flux
calibration obtained from spectra of standard stars. Spatial offsets
between frames were determined from the commanded telescope offsets. These were
found to be repeatable to within an individual pixel (0\farcs043) from the
observations of standard stars, obtained with an identical dither pattern.
Finally, the cube was binned (without any smoothing) in spatial resolution to
match the $0\farcs0996$ WFPC2 pixels, also very close to
the width
of a NIFS slice (0\farcs103). Smoothing was applied in the
spectral dimension, with a Gaussian filter of ${\rm FWHM} = 20$~\AA, well
matched to the final instrumental resolution.

\subsection{Point-Spread Function Comparison}\label{psf_comparison}

The HST and CFHT point-spread functions (PSFs) were characterized from images
of unsaturated stars in the field. For CFHT this was facilitated by
short exposures of the AO guide star interleaved with science observations.
A difficulty for NIFS is that the
$3\arcsec\times3\arcsec$ field of view was too small for this purpose, which
would require
 repositioning of the telescope back and forth between the
science field and the PSF star, thus increasing overhead.
Instead, that PSF was estimated from the non-AO observations
of the telluric calibration star by employing a simple
model of AO correction in which scattered flux is effectively removed 
from the seeing-limited halo and transferred into a diffraction-limited core.

\begin{figure}
\plotone{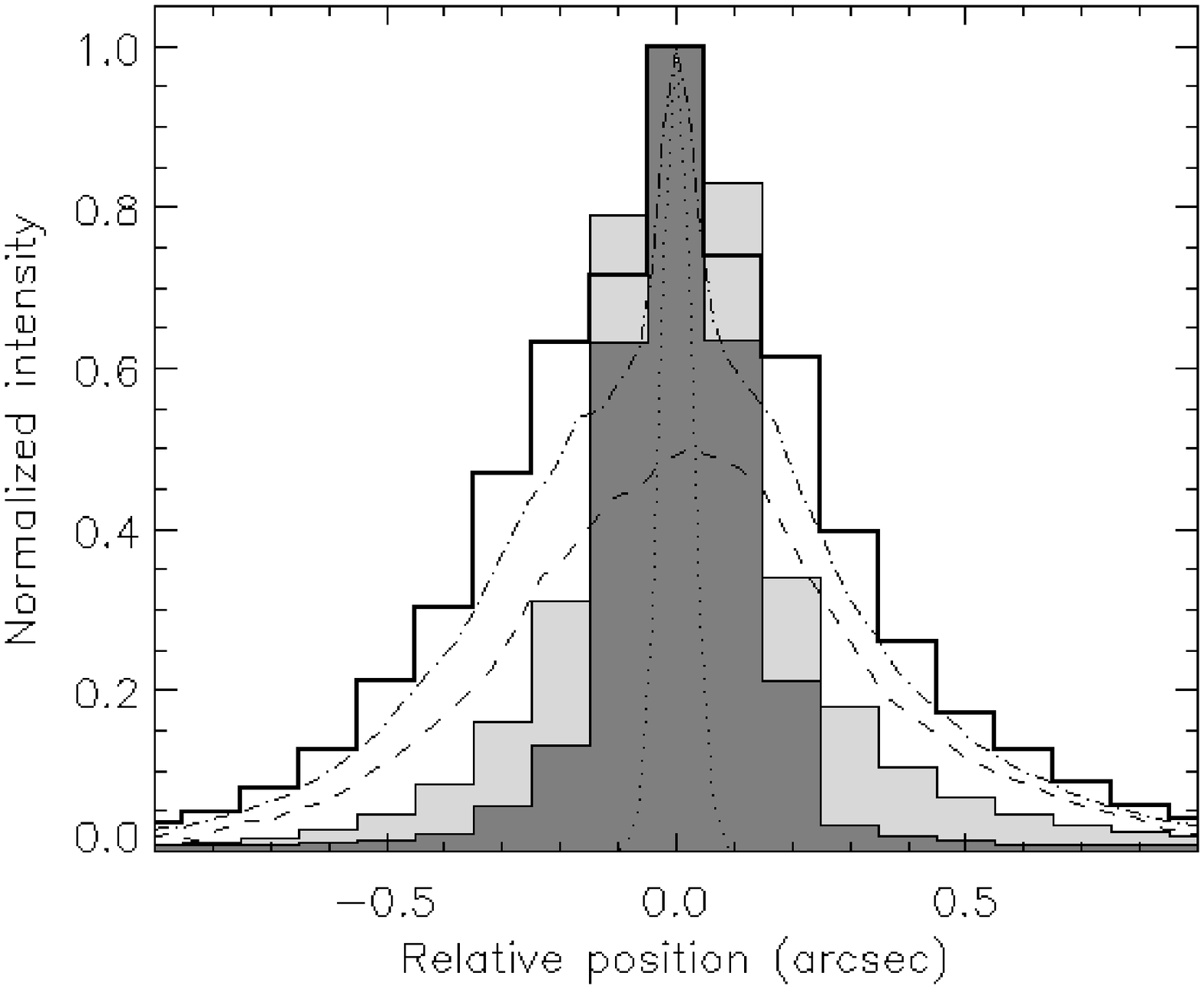}
\caption{The PSF for HST F702W (dark shading) and CFHT AO $K$ (light shading) from
profiles of unsaturated stars, overplotted with an estimate of the Gemini Altair
NIFS PSF profile (thick outline) for Strehl ratio $S=20$\% - all binned to the
HST pixel scale. The NIFS PSF is the combination of an idealized diffraction-limited
core (Gaussian; dotted curve) and a seeing-limited halo
(non-AO PSF; dashed curve); the result
for a Strehl-ratio of $S=35$\% is shown as a dot-dashed curve. Although the resolution of
the NIFS PSF is better than HST and CFHT AO, the
wings are broader than both.}
\label{plot_psf}
\end{figure}

The following method for applying the AO PSF model was used, comparable to that
in \cite{Steinbring2005}. The NIFS
PSF was assumed to have two parts: a diffraction-limited core represented by
a Nyquist-sampled Gaussian distribution, and the observed seeing-limited halo.
These were added in proportion, with the diffraction-limted core given a Strehl
ratio $S$, and the halo receiving the remainder, $1-S$. That the broad 
shape of this halo is not
significantly affected by the AO system is a property of
the finite number of actuators across the telescope pupil. This limits the
greatest angle within which AO can re-direct the light from the halo inwards to
be an ``outer working angle" of $\lambda\over{2d}$, where
$d$ is the physical size of an actuator and $\lambda$ is the observed wavelength.
At $1.65~\mu$m for Altair, with 12 actuators across the pupil giving
$d\approx0.67$m, this angle is 0\farcs25. Outside this angle the AO PSF
is therefore well defined by the $H$-band seeing disk.
The results for $S=20$\% are shown in Figure~\ref{plot_psf} as profiles along
the central NIFS slice centered on the star, and across the peak pixel of
the HST and CFHT PSF (after sub-pixel recentering) all sampled to the HST pixel
scale. For comparison,
the result for $S=35$\% is shown (at the $0\farcs043$ NIFS pixel scale)
which, according to the exposure time calculator available on the Gemini
webpage, would be the best correction possible under 20-percentile seeing.
The two key implications are that the resolution of NIFS imaging
is likely to be as good or better than HST and CFHT, although that is not strongly dependent on
Strehl ratio, and that the wings of the PSF
are somewhat broader. At least 30\% of flux falls
outside a radius of $0\farcs25$, with an upper limit of perhaps 50\%.

So although the NIFS PSF core is narrower than one IFU slice, its 
halo will still lead to cross-talk: contamination in the
spectral dimension of the datacube by light from nearby spatial pixels.
For the faint galaxy spectra discussed here, a 
minimal spatial binning of $0\farcs10$ was found to provide sufficient $S/N$.
Although not implemented, improvement might come from employing PSF
deconvolution. However, a more sophisticated
treatment of the AO system with better
 constraints on its performance - and the seeing - would be
 needed to decouple adjacent spectra, especially those separated by 
 more than $0\farcs10$ and less than $0\farcs25$. It is evident that
care is needed in interpreting the photometry and spectroscopy of the
AO NIFS data on these spatial scales, and that knowledge guided the
analysis to follow.

\subsection{Synthetic Narrow-Band Images}

\begin{figure}
\plotonenarrow{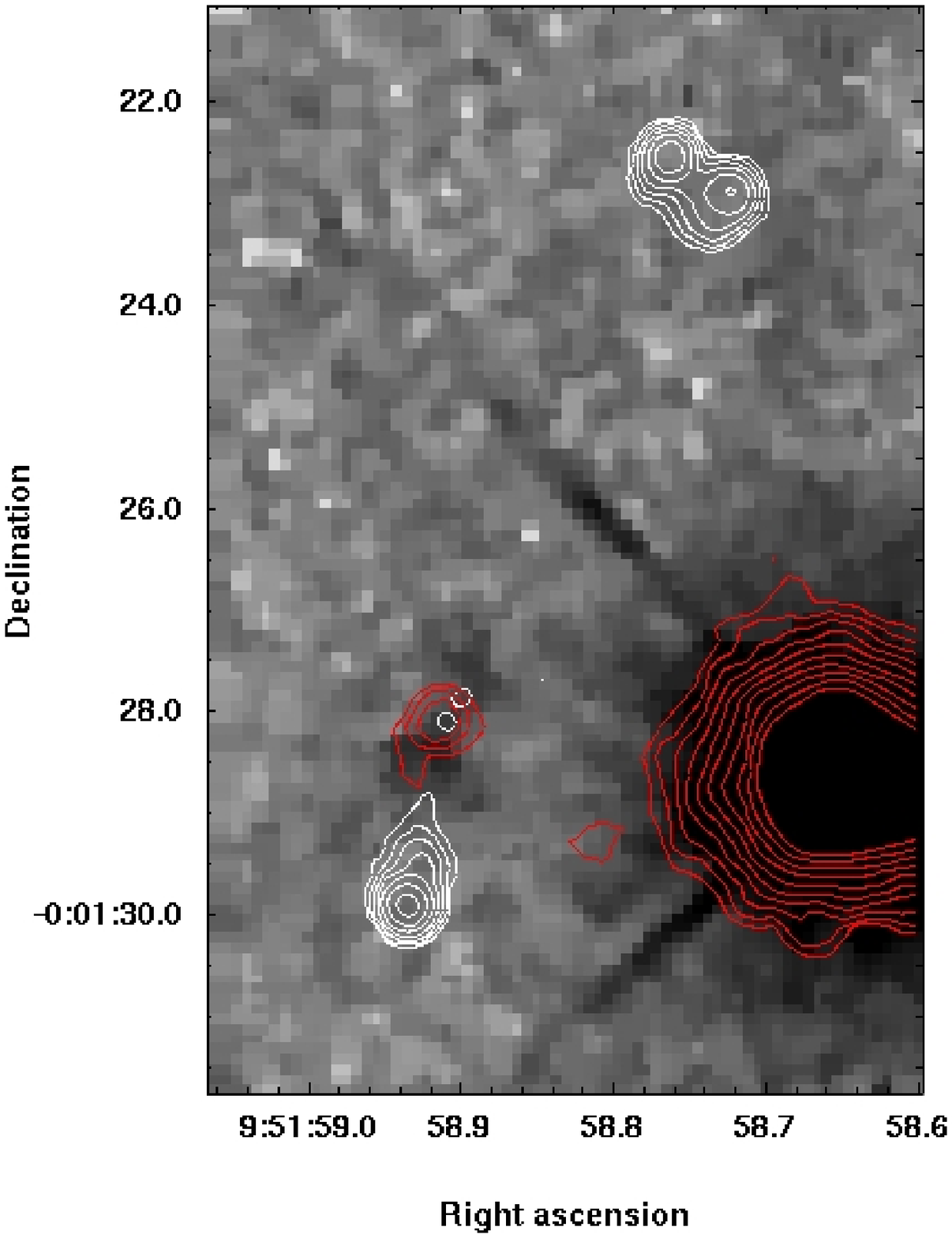}\\
\plotone{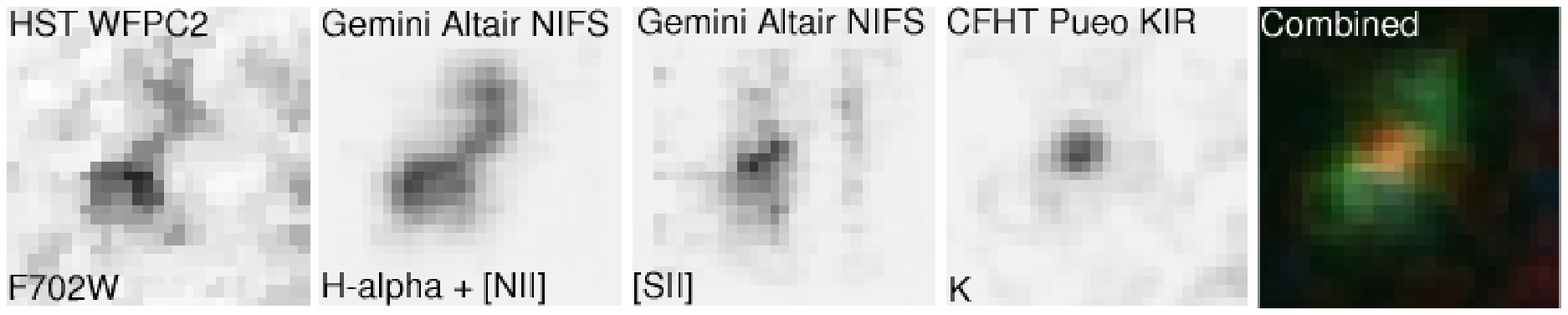}
\caption{The HST WFPC2 F702W image of 3C 230 is overplotted with contours from
the CFHT AO image obtained at K (red) and the VLA radio map at 8.4 GHz (white).
All have been binned to the WFPC2 pixel scale. Below are 2-dimensional
representations of the Gemini-North Altair NIFS data at H$\alpha +$[N {\small
II}] and [S {\small II}] produced by collapsing sections of the data cube along
the spectral dimension, flanked by the HST and CFHT images of the same field
(3\arcsec $\times$ 3\arcsec). Also shown is a ``true-color'' image produced by
combining the HST (blue), Gemini (green), and CFHT (red) images. North is up and
east left.}
\label{plot_finder}
\end{figure}

To begin, synthetic narrow-band images, 30 nm wide, were produced from the NIFS
datacube. These were centered on the only two prominent redshifted lines in $H$-band:
H$\alpha$ (later shown to be blended with [N {\small II}] $\lambda$6548 and
$\lambda$6583) and the [S {\small II}] $\lambda\lambda$6716, 6731 doublet.
These narrow-band images are displayed in Figure~\ref{plot_finder}, along with insets of the
archival HST and CFHT AO images. A ``true-color" image is also shown, using the
combination of H$\alpha+$[N {\small II}] and [S {\small II}] to represent green;
HST, blue and CFHT AO, red. The galaxy is clumpy, running along
a central spine southwest to northeast aligned with the
radio lobes, ending in broader tails with a total span of 2\farcs4. 
The emission-line images indicate
narrow jet-like structures with bends, possibly shock-fronts at 
the edge of an expanding bubble, or edge brightening. 
If due to star formation, then these linear structures
would necessarily be young, as they would not be dynamically stable over
$10^8$ yrs, and there is no evidence from the $K$ image of the galaxy being edge-on.

In general, the UV/optical morphology of 3C 230 exibits the well-known ``alignment effect" \citep{McCarthy1987, Chambers1987} possibly associated with recent star-formation induced by the jet. Another mechanism contributing to the extended component could be 
scattered AGN light, either from electrons or dust. The total flux in $H$ band from CFHT AO is 
$1.48\times 10^{-17}$ erg ${\rm s}^{-1}$ ${\rm cm}^{-2}$ ${\rm \AA}^{-1}$, and 
the sum of Gemini NB images is $1.33\times 10^{-18}$ erg ${\rm s}^{-1}$ ${\rm cm}^{-2}$ ${\rm \AA}^{-1}$, or approximately 11\% of total H-band flux. Based on the composite spectrum of \cite{McCarthy1993} it seems the $H$-band continuum is typical of other HzRGs, for which H$\alpha$ by itself should constitute perhaps 13\% of emission in $H$. Another contributor might be nebular thermal continuum, an extreme case of which is represented by 3C 368 in which only 20\% of the UV continuum could be due to other sources \citep{Stockton1996}.  But the $R-H$ color of 3C 230 suggests that much of its $H$ continuum flux of $1.07\times 10^{-19}$ erg ${\rm s}^{-1}$ ${\rm cm}^{-2}$ ${\rm \AA}^{-1}$ would be from young stars. Of course, for the redshift of 3C 230 F702W falls below the 4000~\AA~break, and so is subject to emission-line contamination, e.g. by Mg {\small II} $\lambda$2800, which may also contribute. This would 
also imply the star-formation is not as young as it appears from its color. Even so, the CFHT $K$ image shown in Figure~\ref{plot_finder} would be expected to better map the true stellar content, and its $H-K\sim0$ color also suggests a contribution by young stars, with a $K$ flux of $5.53\times 10^{-18}$ erg ${\rm s}^{-1}$ ${\rm cm}^{-2}$ ${\rm \AA}^{-1}$ equivalent to 37\% of the $H$-band flux, including $H\alpha$. Thus 3C 230, as with other HzRGs, has an extended UV/optical morphology which may come from a combination of sources, although it is possibly dominated by stellar light in $H$ band.  

\subsection{Spectra}

Next, spectra were extracted from the NIFS datacube. The centroid and width of
both H$\alpha +$[N {\small II}] and [S {\small II}]
were found with two methods: by the FWHM of the peak spectral pixel, and by
attempting multiple Gaussian fits to the line. The results of these methods are
comparable, although the first was more robust, probably because binning the
pixels spatially and already smoothing with a Gaussian kernel spectrally has
improved the $S/N$, with a good match to the instrumental resolution. Choosing a
minimum acceptable $S/N$ of about 50\% of peak flux in the entire image worked
best.

\begin{figure}
\plotonenarrow{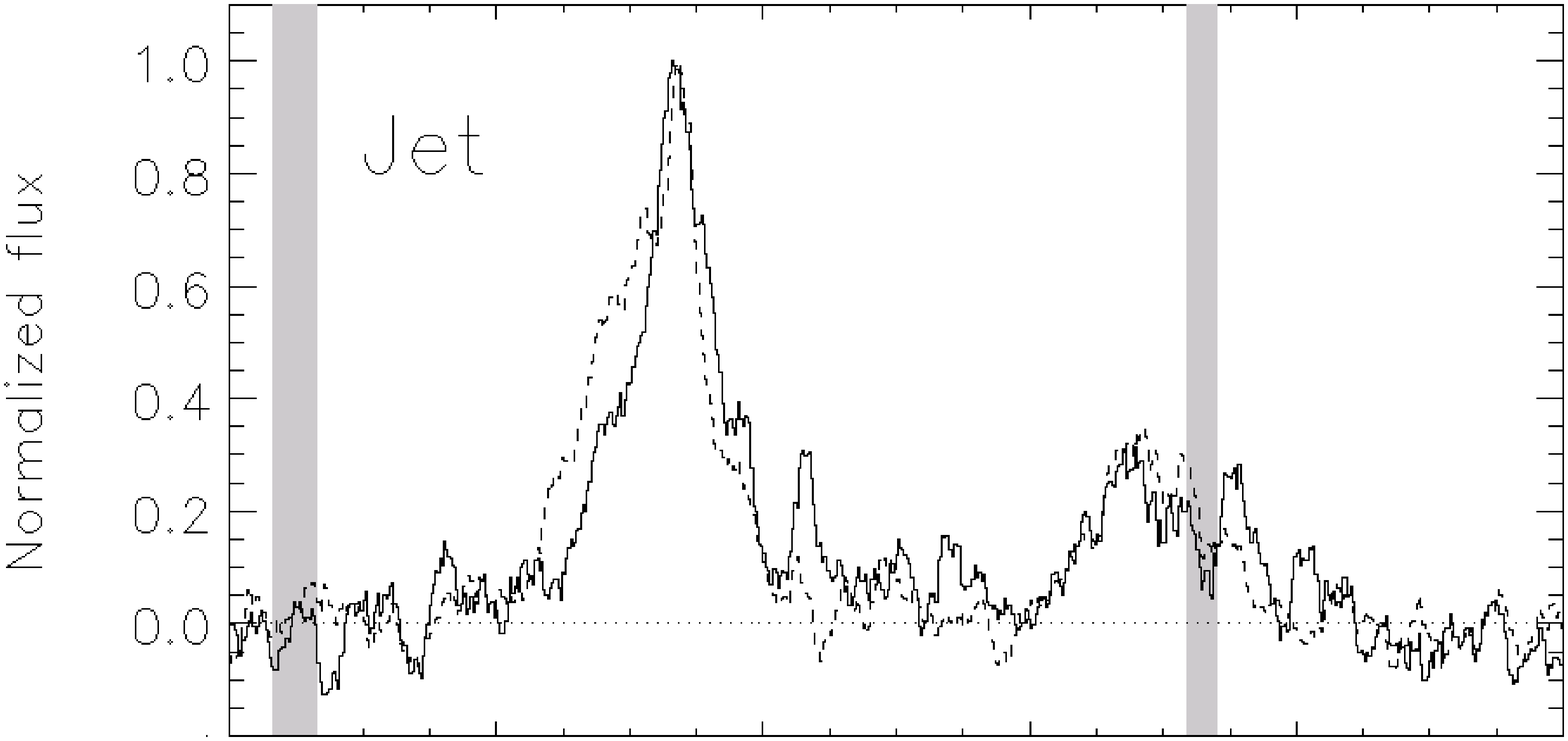}\\
\plotonenarrow{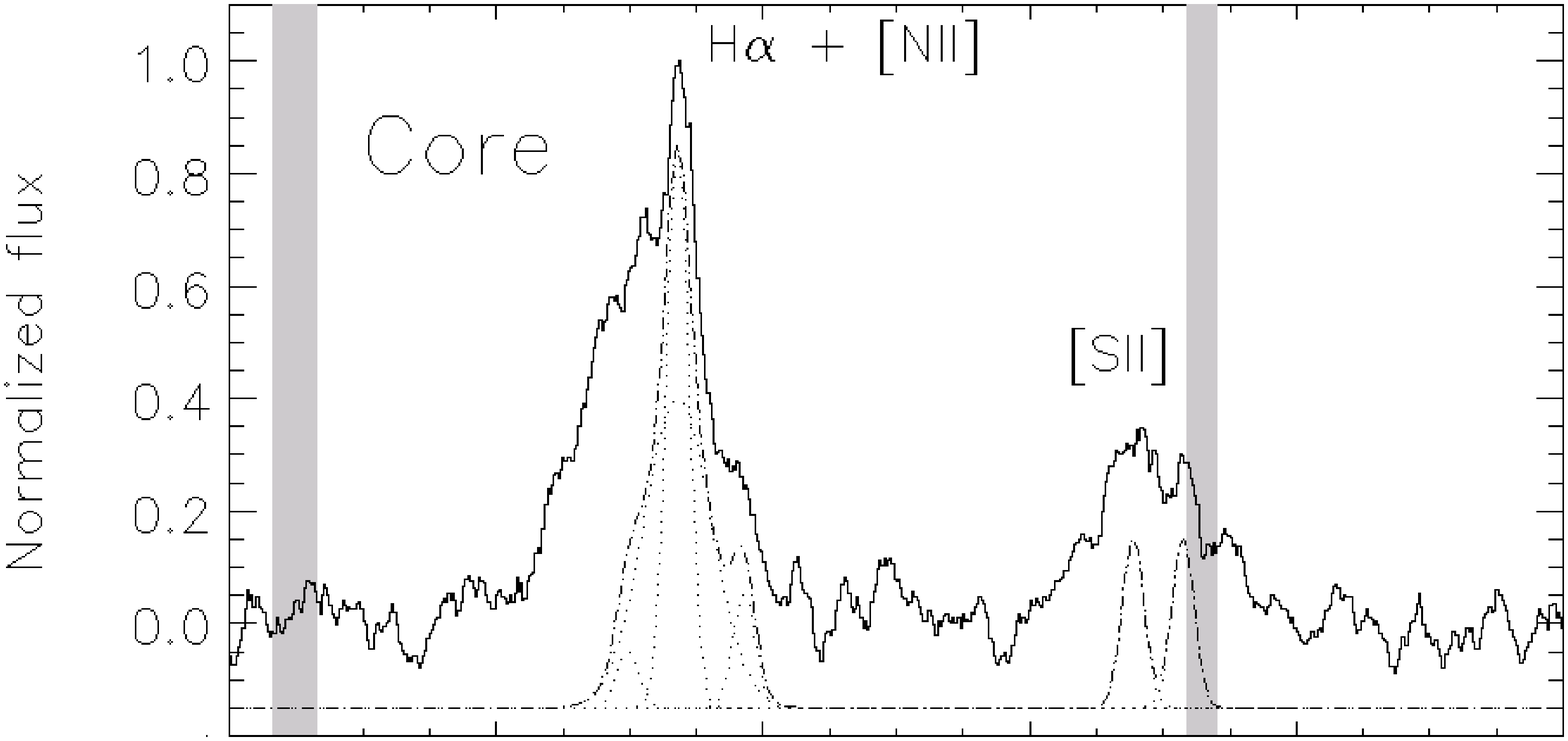}\\
\plotonenarrow{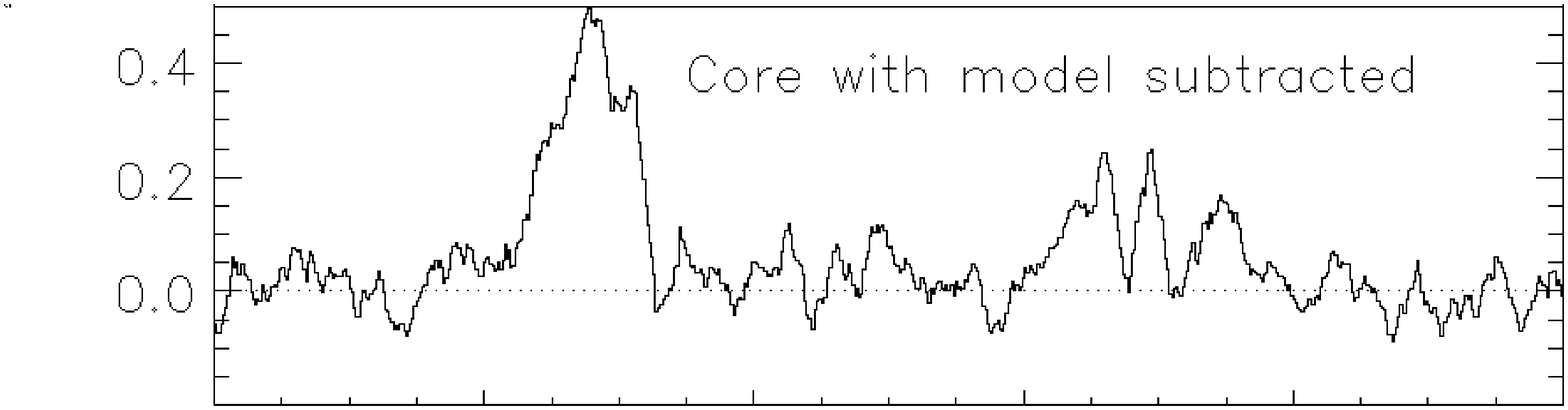}\\
\plotonenarrow{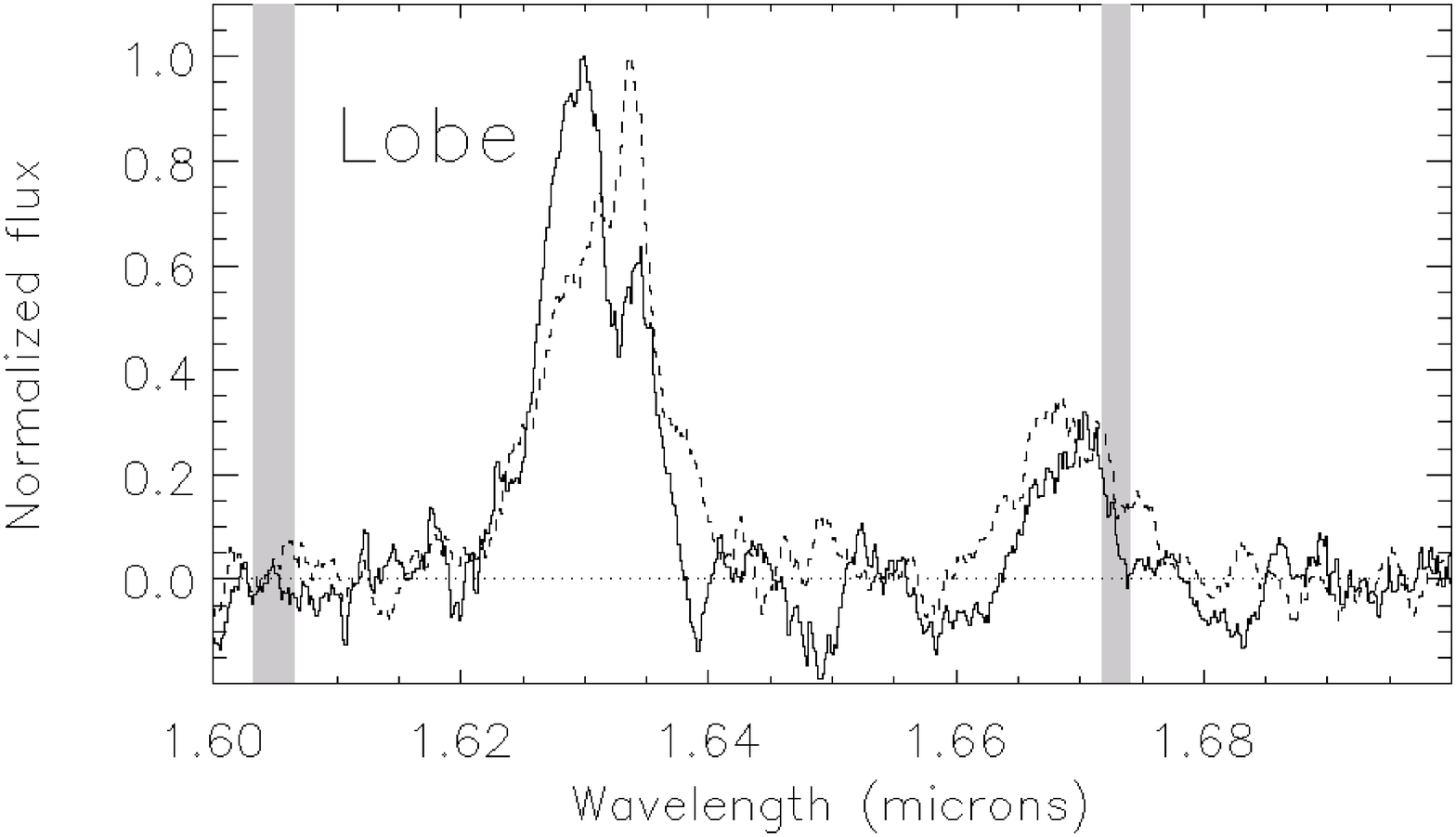}
\caption{Observed-frame spectra of 3C 230 normalized to the peak flux, with 
continuum subtracted. In the top panel is the spectrum of the `Jet', the pixel with
the highest recessional velocity in H$\alpha +$[N {\small II}]. The two panels below this are
the `Core', the central region coincident with
the centroid of the CFHT $K$ image with a model spectrum overplotted
 (dotted curves; dot-dashed envelope), and the residual after subtraction. The bottom panel is the `Lobe', the spectrum at the
peak in H$\alpha +$[N {\small II}] to the southwest. The `Core' spectrum is overplotted
as a dashed curve.}
\label{plot_spectrum}
\end{figure}

Results are shown in Figure~\ref{plot_spectrum}, with continuum subtracted.
The spectrum for the central $3\times3$ pixel
(0\farcs3~$\times$~0\farcs3) region centered on the CFHT $K$ image is labeled the `Core'.
The Gaussian model (dot-dashed curve) is overplotted; a composition of unresolved narrow lines [N {\small II}] $\lambda$6548, $\lambda$6583; and [S {\small II}] $\lambda$6716, $\lambda$6731 and H$\alpha$; plus a broader (120 km/s) H$\alpha$ component (dotted curves) perhaps indicating weak underlying
broadline emission. The mean H$\alpha+$[N {\small II}] centroids gives a systemic galaxy redshift of $z=1.4863\pm0.0002$, corrected to the solar
rest frame (using the IRAF task rvcorrect). A maximal line ratio of [N {\small II}]/H$\alpha=0.45$, determined at
the blended ``shoulder" (providing total luminosity $L_{{\rm
H}\alpha}\approx4\times10^{42}~{\rm erg}~{\rm s}^{-1}$) is in the range typical
of low-redshift RGs and Seyfert 2s \citep{Veilleux1987}, as is the mean [S
{\small II}] to H$\alpha +$[N {\small II}] line-ratio of 0.45 ($\log$[S {\small
II}]/H$\alpha$+[N {\small II}]$=-0.34$), both well above the ``maximal
starburst" limit for ionization attributable solely to H {\small II} regions or
a starburst galaxy \citep[-0.4;][]{Kewley2006}. The latter is comparable to what
\cite{Nesvadba2006} find for the narrow-line region of the $z=2.2$ RG MRC
1138-262, as is a fit of [S~{\small II}]$\lambda 6716$/{\rm [S~{\small
II}]}$\lambda 6731\approx1$ for the approaching gas, although a strong skyline
leads to poor telluric correction in the red tail of the doublet (shaded
region). The residual after subtraction of the model is provided below, demonstrating 
a good fit redward of 1.63 $\mu$m and an unsubtracted component blueward.
Although not shown here, restricting the radius of the `Core' region does not improve
this residual. This is expected, however, due to contamination by the brightest H$\alpha$+[N {\small II}] emission, which is less than 0\farcs5 away (See Section~\ref{psf_comparison}). Spectra at those two individual pixels 
corresponding to the highest receding and approaching H$\alpha$
velocities are displayed in the top and bottom panels, respectively of Figure~\ref{plot_spectrum}. For comparison, the `Core' spectrum is overplotted as a dashed curve. A feature in the receding spectrum (top) is a narrow line at 1.643 $\mu$m, which would correspond to [N {\small II}] $\lambda$6583 receding 920 km ${\rm s}^{-1}$ relative to the systemic galaxy redshift. Possibly this is a secondary, higher velocity component of [N {\small II}] emission along that line of sight, and the weaker part of the doublet is masked.  That interpretation is adopted here, and due to a direct correlation with the radio structure at this spatial pixel (discussed later) this 
position is given the label `Jet'. In the approaching
spectrum (bottom) the shape of H$\alpha+$[N {\small II}] is broadened due to turbulence, and is here labeled `Lobe'.

\subsection{Pixel-by-Pixel Correlations Between All Datasets}

Finally, results are shown in Figure~\ref{plot_contours} for spectra
extracted in every spatial pixel: maps of intensity,
velocity and velocity dispersion - the last after subtracting the instrumental
resolution in quadrature. The intensity maps are overplotted with contours from
HST, CFHT, and VLA. The white hashes indicate the orientation of the radio jet
defined by the alignment of the VLA radio core and the next brightest peak.
Below are shown the (logarithmic) line ratios of H$\alpha+$[N {\small II}] and
[S {\small II}] in each pixel, overplotted with their flux, velocity and
velocity dispersion contours. Note the relative fluxes of [S {\small II}] and
H$\alpha +$[N {\small II}]. Faint [S {\small II}] fluxes at some high
recessional velocities (corresponding to the northeast part of the galaxy) are
due in part to the superposition of a sky line, but could also be
plausibly due to some differential reddening in the galaxy itself, screening
weaker [S {\small II}] along its receding, northern half.

Correlations between the radio map, kinematics, and line ratios are particularly
evident at the three locations discussed above, labeled in
Figure~\ref{plot_contours} as:

\begin{itemize}
\item[{\bf Jet}]\hfill\\
The second brightest VLA radio knot, corresponding to
the highest recessional velocity observed in H$\alpha+$[N {\small II}]. From
here a fainter radio structure extends to the north, comparable to the HST and
Gemini images;
\item[{\bf Core}]\hfill\\
The position of the brightest VLA radio knot, which is coincident with the
kinematic center in the H$\alpha+$[N {\small II}] and [S {\small II}] emission,
and the maximum velocity dispersion in [S {\small II}]. It is also the narrowest
point along the central spine of the HST image, and corresponds to the centroid
of the CFHT $K$ image;
\item[{\bf Lobe}]\hfill\\
The peak in H$\alpha+$[N {\small II}] emission, coincident with both its maximum
approach velocity and velocity dispersion. Notably, [S {\small II}] is weak
here, leading to a trough in [S {\small II}] to H$\alpha$ line ratio. The [S {\small II}] maximum approach
velocity falls a projected distance of $\sim 2$ kpc to the west, closer to a
faint knot of radio emission, just at the detection limit of the VLA map. This
is also the terminus of a ``tail" in the CFHT $K$ image - components `b' and `c'
in \cite{Steinbring2002}.
\end{itemize}

\section{Discussion and Summary}\label{discussion}

A combination of processes are possibly responsible for the strong correlation of 
UV, optical and radio emission in 3C 230: scattered AGN light, nebular thermal 
emission, and past or active star formation. Certainly the radio jet plays a role 
in shaping this morphology. That is consistent with a picture of asymmetric pressure 
and gas content in RG host galaxies \cite[see e.g.][and references therein]{Miley2008}; 
the asymmetry of the 3C 230 radio source is mirrored in the optical surface brightness, 
and in the strong turbulence in the direction of the shorter radio arm to the 
southwest. And in favor of this contributing to star formation is that, despite 
being a powerful radio source, 3C 230 exhibits a relatively low velocity for the bulk 
of its gas. Other HzRGs, for example those discussed by Nesvadba et al., generally have 
higher velocities and line widths.

\begin{figure}
\plotone{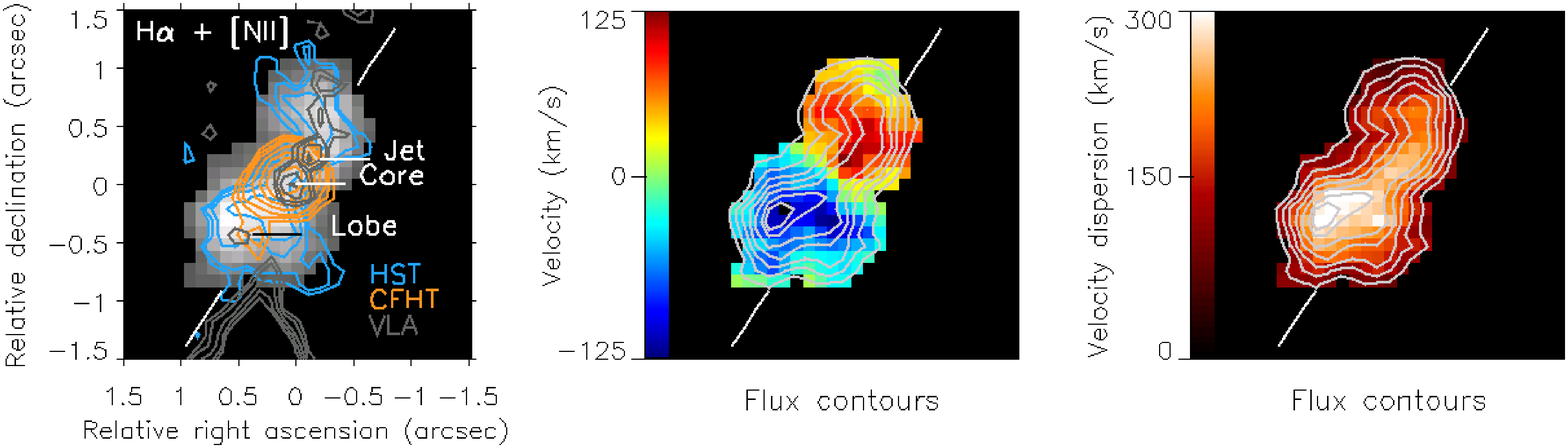}\\
\plotone{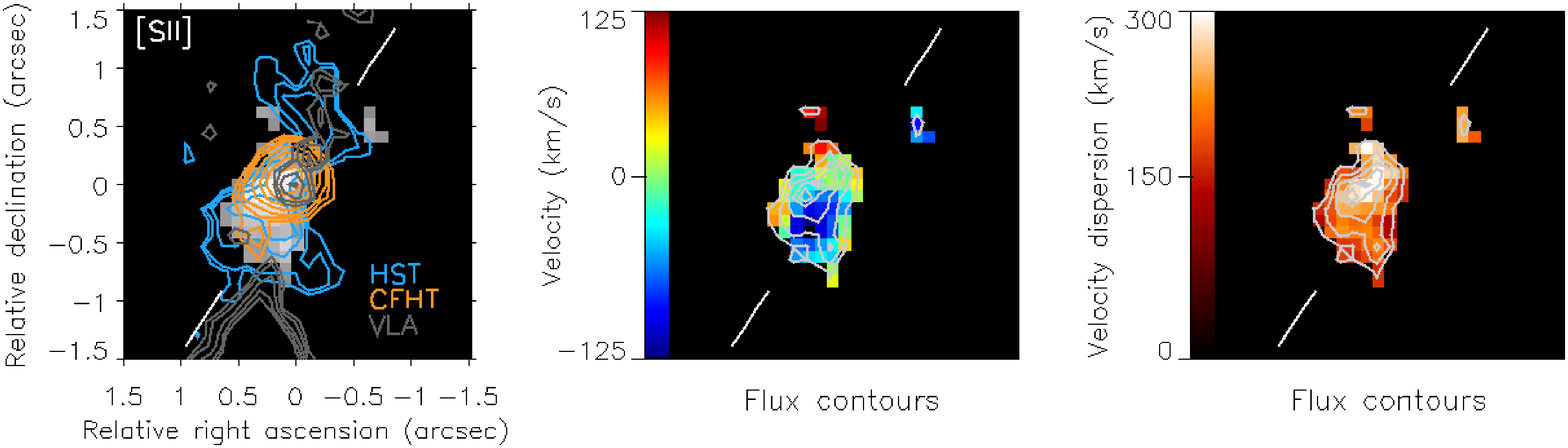}\\
\plotone{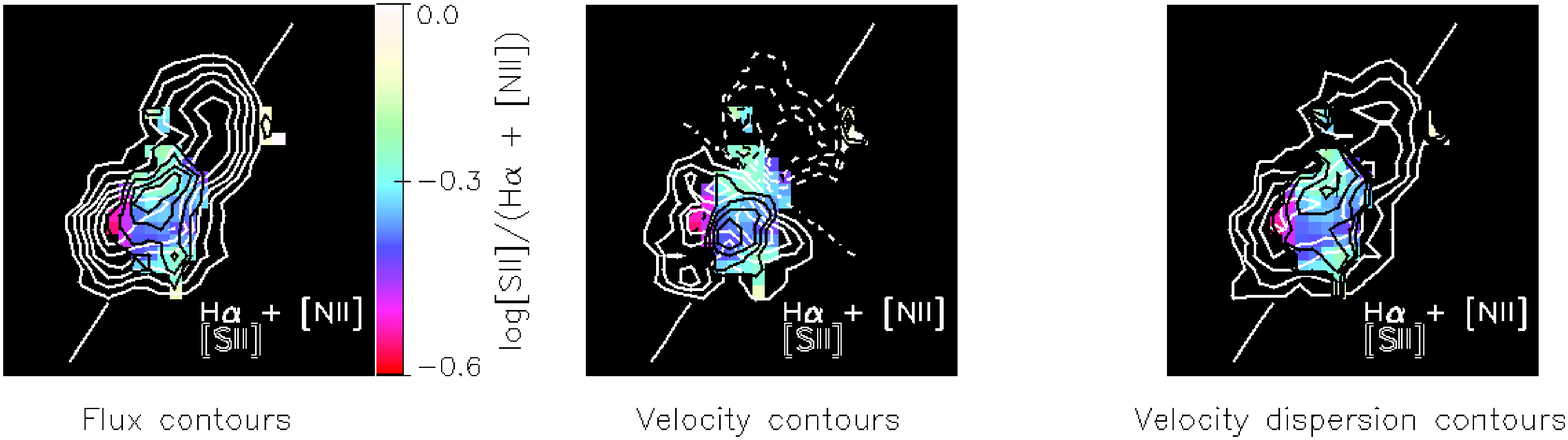}
\caption{Images produced from the H$\alpha+$[N {\small II}] (top left panel) and
[S {\small II}] (middle left) spectra. Overplotted are contour plots of the data
from HST (blue), CFHT (red), and VLA (grey). Bottom left panel is the relative
line ratio of H$\alpha+$[N {\small II}] and [S {\small II}]. The white hashes 
indicate the orientation of the radio jet defined by the alignment of the VLA 
radio core and the next brightest peak. For each, contour plots of the fluxes in 
the Gemini data are overplotted in white. Along the top row are the associated 
velocity field and velocity-dispersion field (overplotted on the line ratio 
images below as contours: white, H$\alpha+$[N {\small II}]; black, [S {\small 
II}]). See text for details.}
\label{plot_contours}
\end{figure}

\begin{figure}
\plotonenarrow{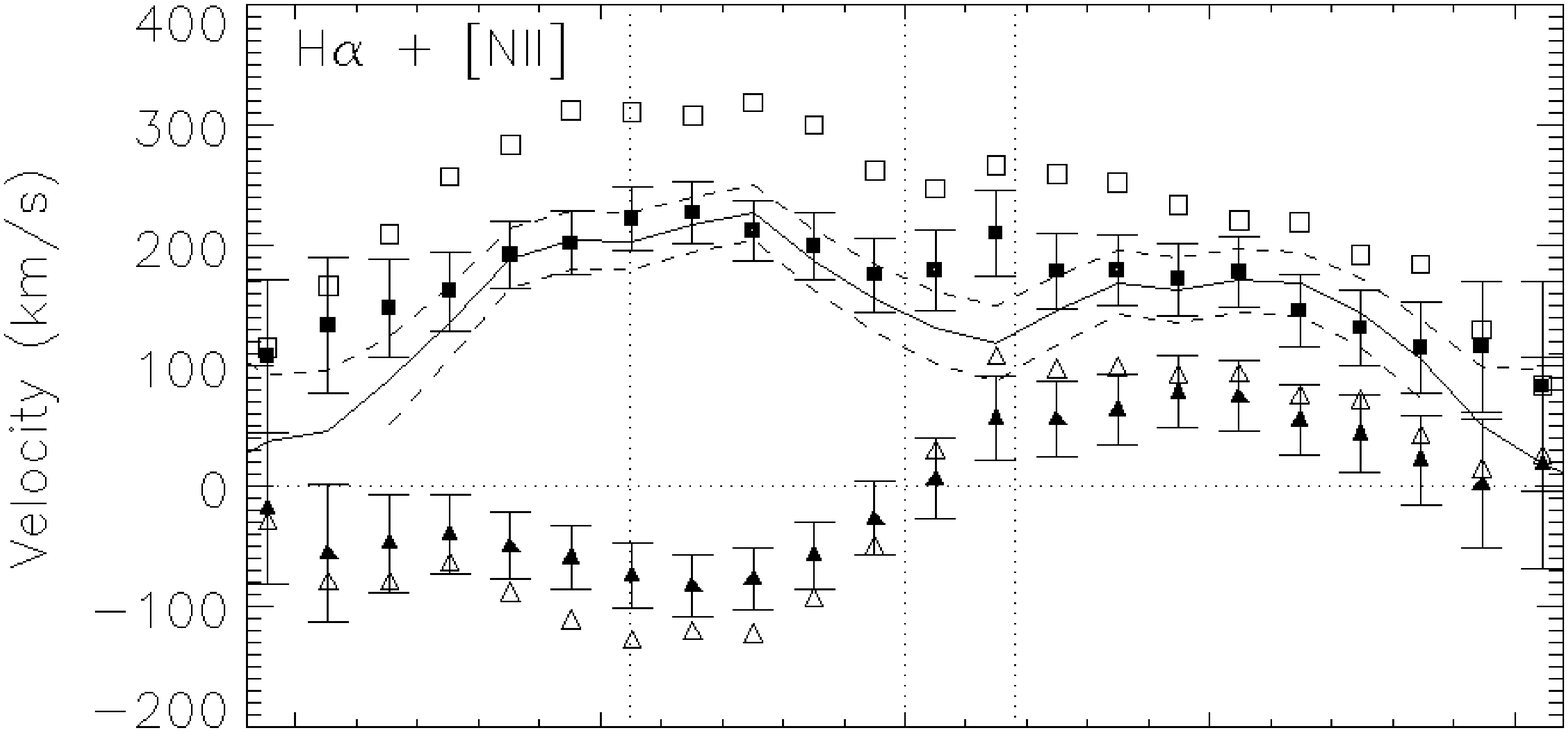}\\
\plotonenarrow{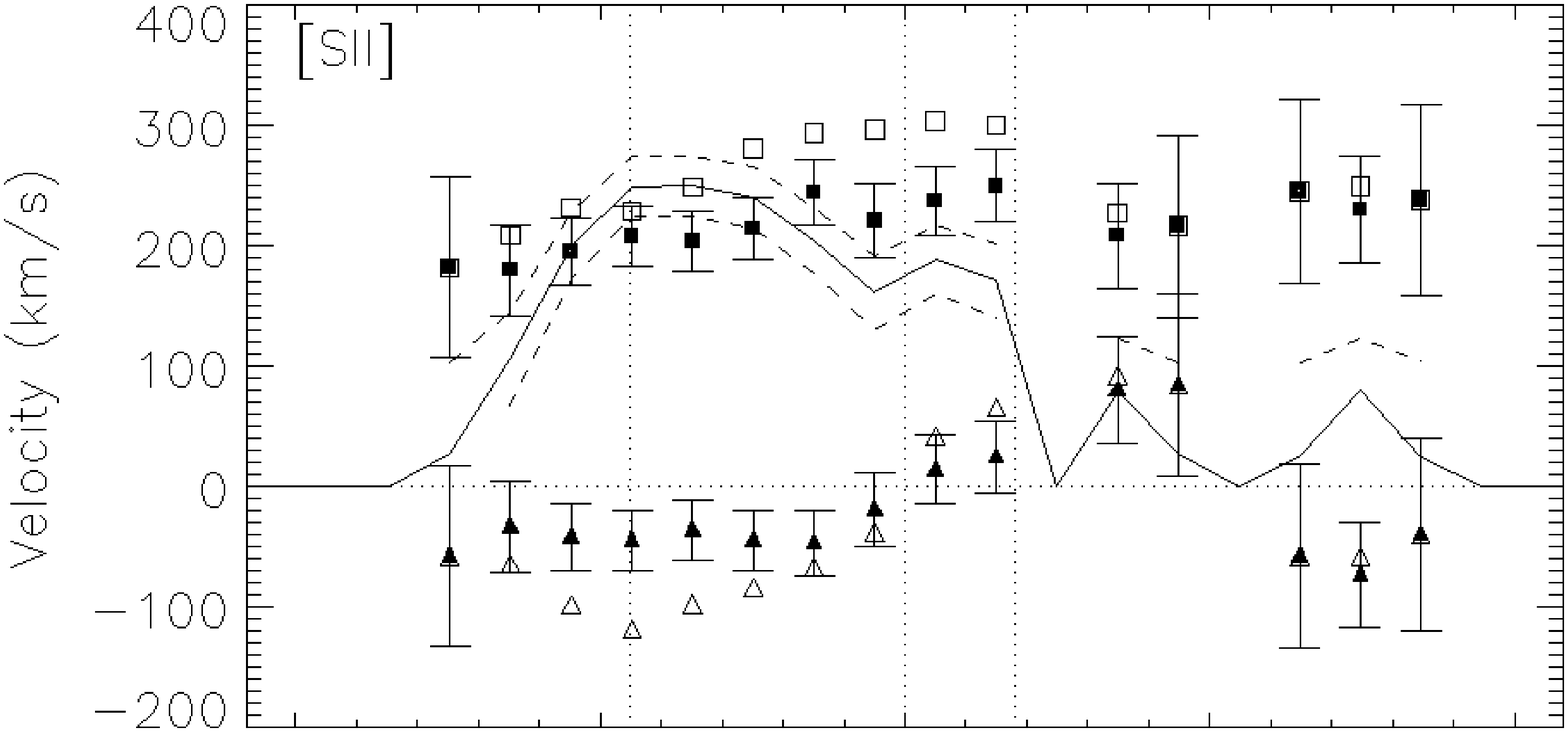}\\
\plotonenarrow{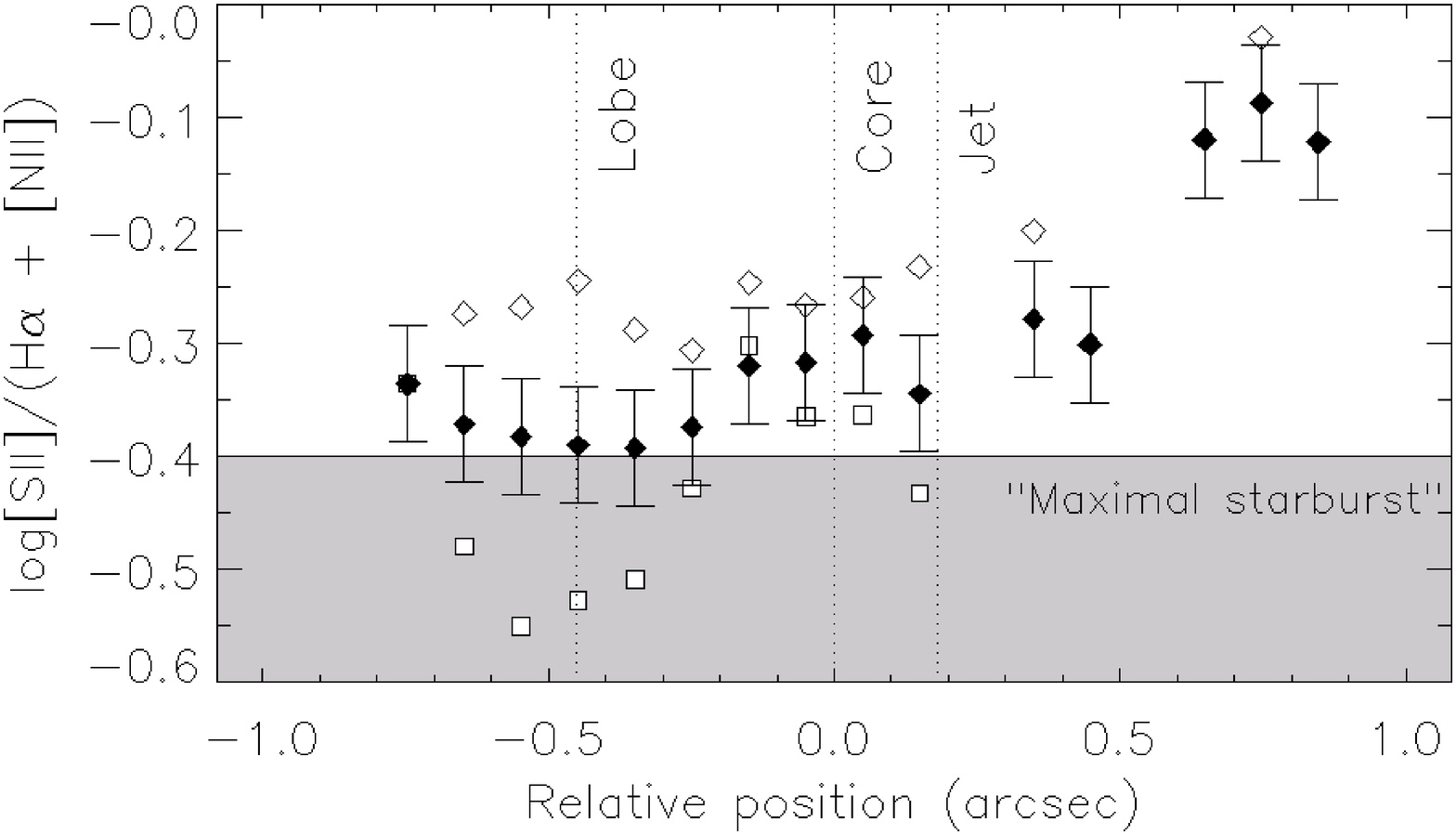}
\caption{Plots of velocity (filled triangles) and velocity dispersion (filled squares) in H$\alpha+$[N {\small II}] and [S {\small II}], averaged transversely 
to the axis of the radio jet; error bars are 1-$\sigma$ deviations, open symbols are maxima. Below are the average relative 
line strengths of [S {\small II}] to H$\alpha+$[N {\small II}] (filled diamonds), maxima (open diamonds), and minima (open squares).}
\label{plot_velocities}
\end{figure}

Transitions in velocity and line ratio between the three key locations along the
radio axis - `Core', `Jet', and `Lobe' - can be interpreted as AGN-driven outflow.
Figure~\ref{plot_velocities} shows the correspondence in more detail. These are
the velocities (triangles) and velocity dispersions (squares) averaged in
one-pixel-wide bins transverse to the radio axis (the white hash-marks in
Figure~\ref{plot_contours}). Error bars indicate 1-$\sigma$ uncertainties at
each position. Open symbols indicate the maxima at each position. Velocity (measured with H$\alpha+$[N {\small II}]) increases sharply from rest to a peak
at the `Jet' location, to the northwest. Note the symmetry of the velocity field
along the radio axis, although less obvious for [S {\small II}] at receding
velocities. In the opposite direction from the `Core', a narrow ridge of bright
emission runs towards the southeast, broadening and smoothly transitioning from
high velocity dispersion in [S {\small II}] to high velocity dispersion in
H$\alpha+$[N {\small II}]. The end of this ridge is the `Lobe' position, a point
where maximum velocity dispersion in H$\alpha$ is reached (open squares, after
subtracting the velocity dispersion in [S {\small II}]) - a stellar clump left
in the wake of the passing jet, or possibly by shocks at the edge of an
expanding bubble.
Note that the position of maximum velocity as seen in [S {\small II}] is not
coincident with H$\alpha+$[N {\small II}] (see Figure~\ref{plot_contours}), the
impression being that the jet (at least the region with higher
[SII]/H$\alpha+$[N {\small II}] line ratio) has moved to the west. Restframe
optical and UV images are consistent with this view: the CFHT $K$ image
(restframe $\sim I$, stellar continuum) has a faint component near the `Lobe'
position, with a bluer knot in the HST F702W (restframe $\sim U$) to the west.

The kinematics are indicative of a bi-polar flow, which to the southeast possibly 
ends in a region of young stars at the `Lobe' position. Although the mass
distribution is clearly elongated, within a radius of $R=$~0\farcs45~$\approx40$
kpc from the position of the `Lobe', a velocity dispersion of $\sigma=200$ km
${\rm s}^{-1}$ and the virial theorem would imply an enclosed mass of
$$M \sim {{5 R \sigma^2}\over{G}} \approx 1.8\times 10^{11}~M_\odot, \eqno(1)$$
($4.8\times10^{11}~M_\odot$ for $R=$1\farcs2, centered at `Core'). The K-band
flux in the `Lobe' is about 20\% of the galaxy total \cite[CFHT AO $K=19.0$
magnitude of `b';][]{Steinbring2002} so just as an order of magnitude estimate
it represents at most a baryonic mass of $1.0\times11^{11}~M_\odot$, for uniform
dark-matter content. This would have
an escape velocity of
$$v_{\rm esc} \sim \sqrt{{2 G M}\over{R}} \approx 630~{\rm km}~{\rm s}^{-1}.
\eqno(2)$$
However, the diluted mass distribution may reduce escape velocities far from the
galaxy center. A simple argument for a lower limit is that since the
approximately 0\farcs45-long distribution is resolved, it is no thinner than 1
pixel, which if that mass ($M_{\rm min}=M/4.5=4\times10^{10}~M_\odot$) were
isolated, would have an escape velocity of $$v_{\rm esc, min} \sim \sqrt{{2 G
M_{\rm min}}\over{R}} \approx 300~{\rm km}~{\rm s}^{-1}. \eqno(3)$$
Again, it should be emphasized that these are just order of magnitude estimates.
But to further illustrate that this provides a plausible picture, overplotted in
Figure~\ref{plot_velocities} is the average flux along the radio axis (solid
curve, with an arbitrary normalization) and its relative photometric uncertainty
(dashed, assuming only Poisson noise). The mean velocity dispersion in H$\alpha$
is well matched by a mass-follows-light estimate of the luminosity, with peak
velocities in the turbulent regions approaching $v_{\rm esc, min}$, but still
well below the escape velocity of the galaxy.

It is interesting to calculate a limit to the bulk flow of gas, and compare that
to a potential star formation rate in 3C 230. Carrying on with simple
order-of-magnitude estimates, neglecting any contribution from (isotropic)
stellar winds, and assuming a continuous maximal jet velocity of $v=127~{\rm
km}~{\rm s}^{-1}$, the flow has been underway for
$$\tau \sim R/v \approx 3 \times 10^7~{\rm years}, \eqno(4)$$
comparable to the expected duration of AGN activity based on the separation of
the
radio lobes. Following closely the analysis in \cite{Nesvadba2006} of MRC
1138-262, also with a ratio of [S{\small II}]$\lambda 6713$/[S {\small
II}]$\lambda 6731\approx1$, 3C 230 may exhibit a similar ``partially ionized
zone" case, with electron densities of $n_{\rm e}\approx 100-500~{\rm cm}^{-3}$
and temperature $T\sim10^4$ K. A luminosity of ${\rm L}({\rm H}\alpha)=4\times
10^{42}$ erg ${\rm s}^{-1}$, all from recombination-line flux of hydrogen, might
be at most a gas mass of \citep[e.g.,][case-B recombination]{Osterbrock1989}
$$M_{\rm H}\sim 9.73\times10^8~L_{{\rm H}\alpha, 43}~n_{\rm e, 100}^{-1}\approx
4\times10^8 M_\odot, \eqno(5)$$
for $L_{{\rm H}\alpha, 43}$ in units of $10^{43}~{\rm erg}~{\rm s}^{-1}$ and
$n_{\rm e, 100}$ in units of $100~{\rm cm}^{-3}$. This is roughly $4\times10^8
M_\odot/1\times10^{11} M_\odot<1\%$ of the stellar mass in the `Lobe' 
\citep[overestimated in][prior to reducing the doublet]{Steinbring2010}. The 
flow is too slow to escape the galaxy entirely, instead redistributing gas 
during the period of jet activity, and if continuously, at a rate of
$${M_{\rm H}/\tau} \approx 14~M_\odot~{\rm yr}^{-1}. \eqno(6)$$
Of course, this would be for all H$\alpha$ flux coming from AGN-ionized gas. On 
the other hand, assuming the \cite{Kennicutt1994} relation for star-formation 
from H$\alpha$ luminosity, if it were all from young stars this might be at most
$${\rm SFR} \sim 44.4~L_{{\rm H}\alpha, 43}\approx 18~M_\odot~{\rm yr}^{-1}.
\eqno(7)$$
Bringing the argument full circle, this represents at most
$18~M_\odot~{\rm yr}^{-1}\times\tau = 4\times10^9~M_\odot$,
less than 1\% of the galaxy mass. That the rates in
equations 6 and 7 are comparable does not imply a balance
between the two H$\alpha$-emission/ionization scenarios (they are both maxima, 
and so mutually exclusive) although that would be intriguing. But it may at 
least hint at the close relationship between AGN outflow and star-formation - a 
mixture of ionization states also being consistent with the trend in line-ratios 
along the jet axis of 3C 230. 

In conclusion, star formation may be on-going in 3C 230 but it does not constitute 
a significant fraction of the galaxy mass. And as 3C 230 is among those RGs at the 
bright edge of the $K-z$ relation, and already massive, perhaps this is an
example of how AGNs regulate growth: balancing outflow with modest star formation.
Further laser-AO spectroscopy in $J$-band - 
allowing an additional line-ratio of [O {\small III}]$\lambda 
5007$/H$\beta\lambda 4861$ - would help differentiate between ionization 
scenarios on a pixel-by-pixel basis near the AGN, resolving some of the 
ambiguity in flow-rate versus stellar content, and star-formation history. 
Spectropolarimetric measurements could also help in this regard by
quantifying the fraction of light scattered into the observed line of sight.
This study of 3C 230 is a demonstration, a first attempt at laser-AO 
2-dimensional spectroscopy of a high-$z$ radio galaxy where existing HST and 
radio maps are available at comparable spatial resolution. For Gemini Altair 
NIFS, five other such targets are possible, and data for one of those now in 
hand (4C+41.17) will be discussed in a later paper.

\acknowledgements

Teddy Cheung kindly provided the VLA A-array image, and position of the core. I 
would also like to thank the staff of Gemini observatory, especially observers 
Andy Stephens and Thomas Dall. This work is based in part on observations 
obtained at the Gemini Observatory, which is operated by the Association of 
Universities for Research in Astronomy, Inc., under a cooperative agreement with 
the NSF on behalf of the Gemini partnership: the National Science Foundation 
(United States), the Science and Technology Facilities Council (United Kingdom), 
the National Research Council (Canada), CONICYT (Chile), the Australian Research 
Council (Australia), Ministério da Ciência e Tecnologia (Brazil) and Ministerio 
de Ciencia, Tecnología e Innovación Productiva (Argentina).

\newpage

\newpage

\newpage

\newpage

\newpage

\end{document}